\documentclass[sigconf]{acmart}

\usepackage{graphicx}
\usepackage{caption}
\usepackage{subcaption}
\usepackage{pifont}

\AtBeginDocument{%
  }

\setcopyright{acmlicensed}
\copyrightyear{2024}
\acmYear{2024}
\acmDOI{XXXXXXX.XXXXXXX}

\newcommand{\cmark}{\ding{51}}%
\newcommand{\xmark}{\ding{55}}%

\acmConference[CIKM '24]{The Conference on Information and Knowledge Management}{Oct 21--25,
  2024}{Boise, ID}
\acmBooktitle{3rd International Workshop on Industrial Recommendation Systems (at
CIKM 2024).}
\acmISBN{978-1-4503-XXXX-X/18/06}

\begin{document}

\title{Improving feature interactions at Pinterest under industry constraints}

\author{Siddarth Malreddy}
\email{smalreddy@pinterest.com}
\affiliation{%
  \institution{Pinterest}
  \city{San Francisco}
  \state{CA}
  \country{USA}
}

\author{Matthew Lawhon}
\email{mlawhon@pinterest.com}
\affiliation{%
  \institution{Pinterest}
  \city{New York}
  \state{NY}
  \country{USA}
}

\author{Usha Amrutha Nookala}
\email{unookala@pinterest.com}
\affiliation{%
  \institution{Pinterest}
  \city{San Francisco}
  \state{CA}
  \country{USA}
}

\author{Aditya Mantha}
\email{amantha@pinterest.com}
\affiliation{%
  \institution{Pinterest}
  \city{San Francisco}
  \state{CA}
  \country{USA}
}

\author{Dhruvil Deven Badani}
\email{dbadani@pinterest.com}
\affiliation{%
  \institution{Pinterest}
  \city{San Francisco}
  \state{CA}
  \country{USA}
}



\renewcommand{\shortauthors}{Malreddy et al.}

\begin{abstract}
Adopting advances in recommendation systems is often challenging in industrial settings due to unique constraints. This paper aims to highlight these constraints through the lens of feature interactions. Feature interactions are critical for accurately predicting user behavior in recommendation systems and online advertising. Despite numerous novel techniques showing superior performance on benchmark datasets like Criteo, their direct application in industrial settings is hindered by constraints such as model latency, GPU memory limitations and model reproducibility. In this paper, we share our learnings from improving feature interactions in Pinterest's Homefeed ranking model under such constraints. We provide details about the specific challenges encountered, the strategies employed to address them, and the trade-offs made to balance performance with practical limitations. Additionally, we present a set of learning experiments that help guide the feature interaction architecture selection. We believe these insights will be useful for engineers who are interested in improving their model through better feature interaction learning.
\end{abstract}




\maketitle

\section{Introduction}
\begin{figure}
  \centering
  \includegraphics[width=\linewidth]{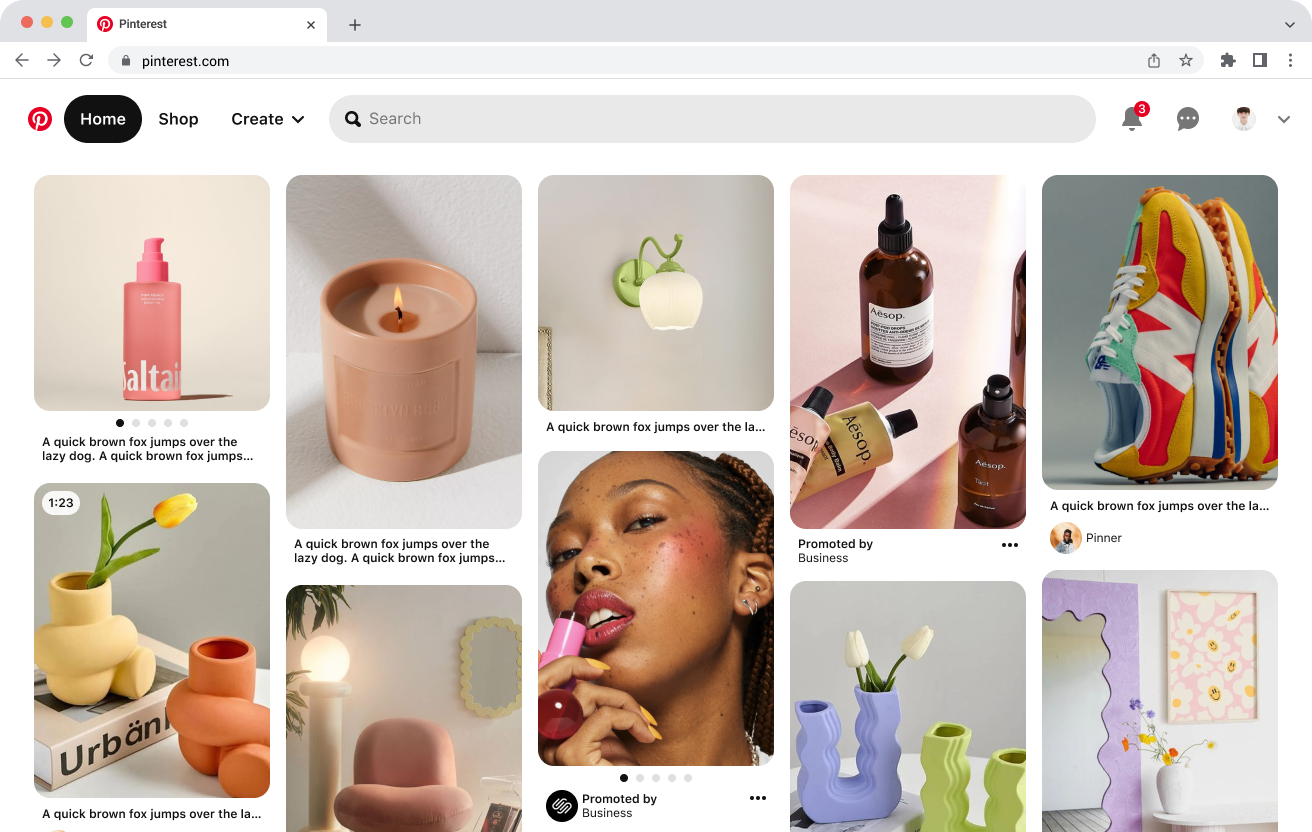}
  \caption{Pinterest Homefeed Page}
  \Description{Pinterest Homefeed Page}
  \label{fig:homefeed}
\end{figure}

Pinterest is one of the largest content sharing platforms with over 500M monthly active users\cite{pint}. The Homefeed, shown in Figure \ref{fig:homefeed}, serves as the primary entry point for most users and is a major source of inspiration, accounting for the majority of the user engagement on the platform. Users on Pinterest can perform a variety of actions such as save, close-up, hide, etc. to interact with the Pins. To enhance our users’ experience on Homefeed, we use a recommendation system to ensure we serve the most relevant Pins to any given user. We use a standard retrieval, ranking and blending based recommendation system. The ranking model is responsible for predicting the probability of different user actions. This is achieved using a multi-task modeling approach.

\begin{figure*}[hbt!]
\centering
\begin{subfigure}{.5\textwidth}
  \centering
  \includegraphics[width=.7\linewidth]{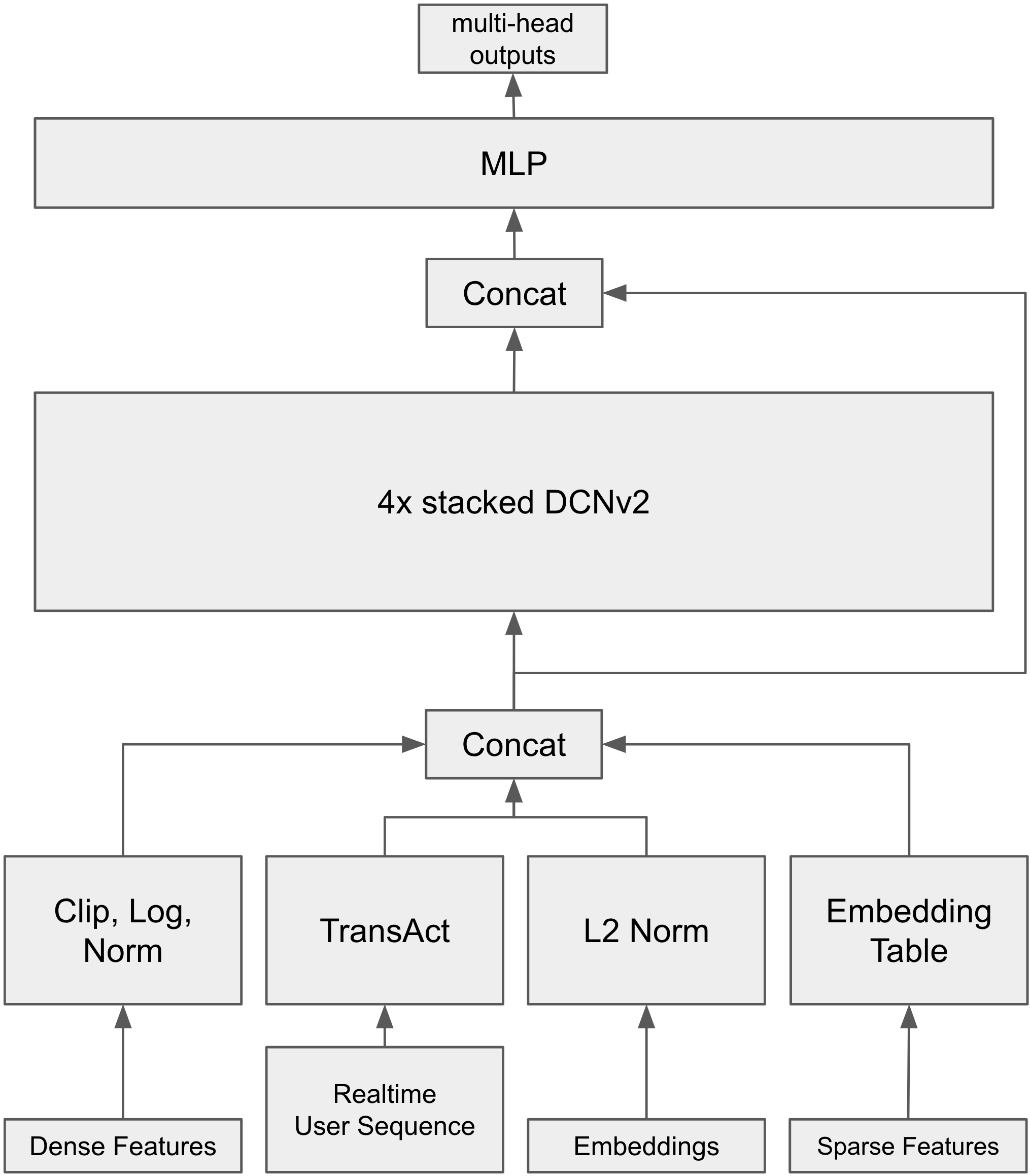}
  \caption{Baseline Configuration}
  \label{fig:dcn}
\end{subfigure}%
\begin{subfigure}{.5\textwidth}
  \centering
  \includegraphics[width=.7\linewidth]{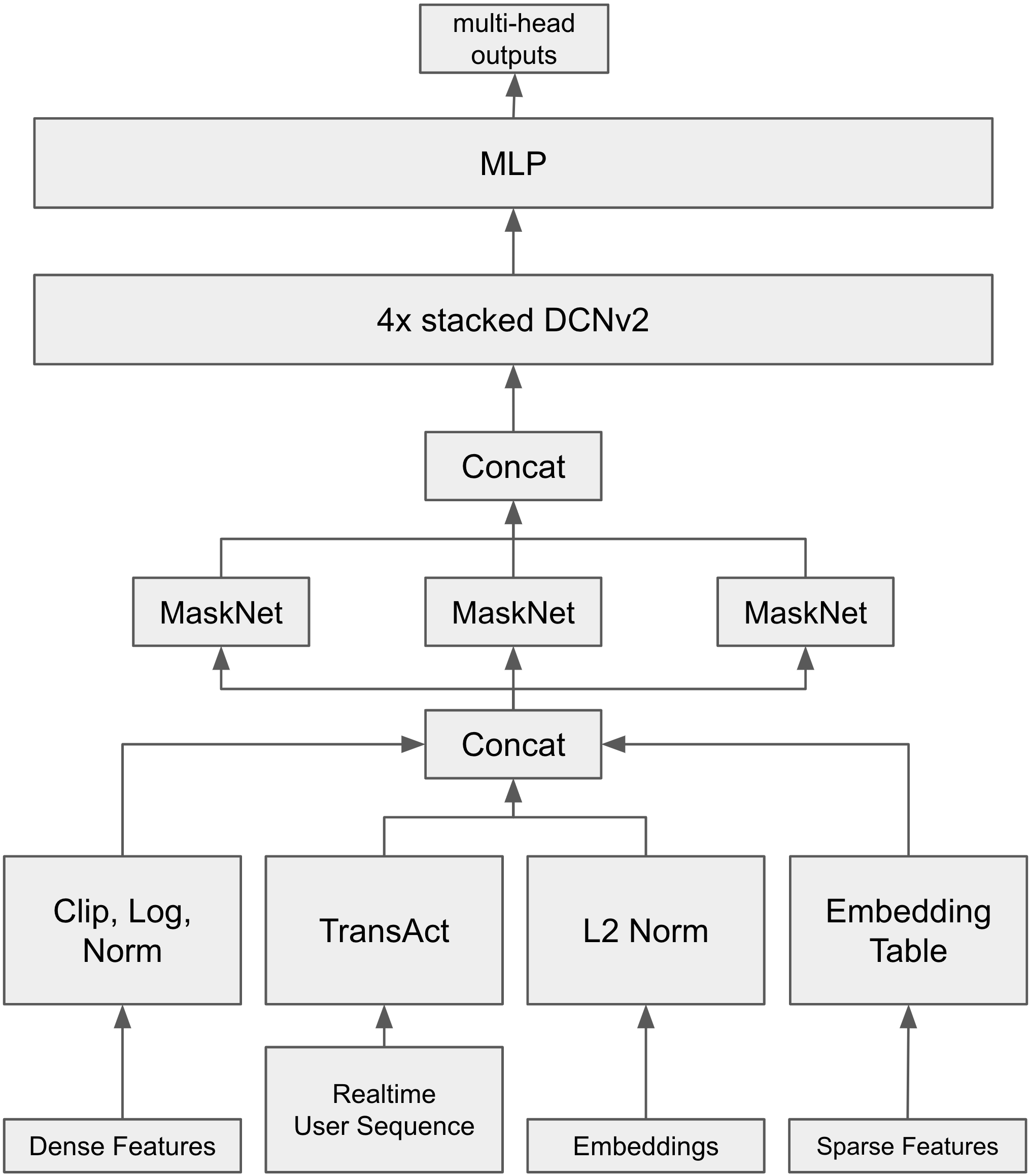}
  \caption{New Configuration}
  \label{fig:masknet}
\end{subfigure}
\caption{Homefeed ranking model}
\label{fig:test}
\end{figure*}

Our ranking model can be categorized into three parts: feature preprocessing, feature interaction, and task prediction. Feature interaction is a critical part of the model, essential for effectively capturing the complex relationships between features and labels. In recent years\cite{gdcn}\cite{masknet}\cite{finalmlp}\cite{memonet}, many architectures have been proposed to efficiently learn these relationships. However, these architectures are often evaluated in an offline setting on benchmark datasets and do not consider the real-world constraints that industrial recommendation systems face. Limitations such as excessive number of hyperparameters and high memory utilization make these techniques impractical in industrial settings. Navigating these constraints is key to improving our model and providing value to our users.

The core contributions of this paper are summarized as follows:
\begin{itemize}
\item We detail the constraints faced in industrial settings and contrast them with academic research.
\item We present a set of learning experiments to help guide the selection of appropriate interaction architectures and hyperparameters.
\item We share our learnings about improving feature interactions in the ranking model using the learning experiments under these industry specific constraints.
\end{itemize}

The remainder of this paper is organized as follows: Section 2 reviews related work. Section 3 describes the Homefeed ranking model. The constraints we work with are detailed in Section 4. Our learning experiments and experiment results using different feature interaction layers are reported in Section 5. Finally, we conclude our work in Section 6.

\section{Related Work}
Most industrial recommendation systems today are based on deep neural network models and a key part for most of these models is to effectively learn the feature interactions. Architectures like the Wide \& Deep\cite{widendeep} and DeepFM\cite{deepfm} show that learning lower order and higher order feature interactions is useful. \cite{paperswithcode} shows that Click-Through Rate Prediction on the Criteo dataset has been getting better over the years with better feature interaction architectures.

An MLP allows for implicitly modeling higher order feature interactions. DCN\cite{dcn} and its successor DCNv2\cite{dcnv2} add an explicit feature cross to learn better interactions. Similarly, xDeepFM\cite{xdeepfm} uses a Compressed Interaction Network (CIN) to learn the lower order and higher order feature interactions in an explicit way. AutoInt\cite{autoint} proposes an attention based mechanism to explicitly model the feature interactions in a low-dimensional space. 

Masknet\cite{masknet} uses an instance-guided mask to aid the interaction learning. FinalMLP\cite{finalmlp} shows a well-tuned two stream MLP model can outperform explicit crosses. GDCN\cite{gdcn} uses gating to filter out noisy feature interactions allowing for even higher-order feature interactions. DHEN\cite{dhen} proposes a framework combining multiple feature interactions together. SDCNv3\cite{li2024dcnv3generationdeepcross} proposes to use a Shallow \& Deep Cross Network which integrates both low-order and high-order feature interactions. Despite the increased accuracy from these new architectures, they increase the latency and memory consumption of the model which makes it harder to use in industry settings.

DeepLight\cite{deeplight} proposed to alleviate this problem by using a mechanism to prune the neural network to reduce inference time while maintaining the model performance. However, pruning introduces the risk of losing reproducibility, which is important in industry settings. The transformer architecture, which uses a multi-head attention mechanism to be able to summarize a set of tokens, has been shown to be effective at learning feature interactions\cite{hiformer}.

\section{Homefeed Ranking Model}

The Homefeed ranking model is a deep learning model responsible for predicting point-wise estimates of user engagement probabilities. Given a user $u$ and a Pin $p$, the model predicts $K$ probabilities - one for each of the $K$ user actions we care about like save, close-up, hide, etc. We use a combination of dense features, sparse features and embedding features to represent the user, the Pin and the context. The dense features are normalized for numerical stability. We represent the sparse features using learnable embeddings and select the embedding size depending on the cardinality of the feature. We project larger embedding features onto a smaller dimension before feeding them into the model. One of the most important embedding features is our user sequence embedding. It is learned with a transformer based architecture\cite{transact} using the user's past engagements as input. The output from this transformer is pooled into an embedding feature. The sparse and embedding features are L2 normalized before being concatenated with the dense features into a single feature embedding. This feature embedding is used as input to the feature interaction layers.
We use 4 stacked full-rank DCNv2\cite{dcnv2} layers to model the feature interactions. We concatenate the crossed feature embedding and the input feature embedding before passing it into the MLP layer. We use a shared MLP with multiple hidden layers and predict $K$ outputs corresponding to the $K$ tasks. The model is trained on users' past engagement data using weighted cross entropy loss. The weights in the loss are selected according to business needs. We treat this architecture shown in Figure \ref{fig:dcn} as the baseline in all of our experiments.


\section{Constraints}
When iterating on an industrial recommendation system, it is crucial to account for the various constraints on the ranking model. Below, we outline some of these constraints, focusing specifically on those pertinent to our feature interaction experiments, and contrast them with academia. This is not an exhaustive list, as it excludes other important considerations such as diversity and balancing business metrics.

\subsection{Memory}
To effectively utilize our computational resources, the batch size used to train our model is chosen so that the maximum allocated memory during model training is approximately 60\% of the total available memory. This allocation leaves room for system memory, memory fragmentation, and potential future projects that might increase memory usage. Though we observe that increasing this further improves our model quality, if the maximum allocated memory exceeds 75\%, we encounter intermittent out-of-memory (OOM) errors during model retraining. To mitigate this, we can reduce the batch size but it comes at the cost of model quality and introduces a confounding variable in our experimentation. Therefore, any new technique that requires substantial additional memory must significantly improve model quality to offset the loss due to the reduced batch size.

This is usually not a consideration for academic research where the batch size may be tuned to maximize the target metrics and training time.

\subsection{Latency}
Model inference latency is a significant component of our overall system latency. Any increase in this latency reduces the throughput of our distributed serving system. As a result, we then need to add additional machines to get the same serving throughput leading to higher operational costs. It is imperative to ensure that any improvements in the model justify these additional costs. Furthermore, increased model latency can extend the training time, thereby impacting development velocity.

A 5\% increase in latency may not have critical effects in academic research but in industrial application it would increase serving costs significantly. That being said, it is one of the easier constraints to trade-off because of easy access to additional compute.

\subsection{Hyperparameters}
Using a model architecture with numerous hyperparameters increases the number of models that need to be trained to identify the optimal configuration. 
As our users' interests change, our data distribution changes. This requires us to tune these hyperparameters at a regular cadence to optimize the model for the latest data distribution.


Academic papers usually perform extensive hyperparameter tuning to identify the best performing variant. Since it is a one-time cost, reducing the number of hyperparameters in the model usually is not a consideration.

\subsection{Reproducibility}
It is crucial to ensure that our model can be re-trained using the same data and produce consistent results. Without this consistency, it becomes difficult to determine whether observed metric movements are genuine improvements for a given change or simply variations within the model. Therefore, any feature interaction incorporated into the model should not decrease its reproducibility.

We measure reproducibility by computing the standard deviation of the HIT@3/save metric amongst the runs using the same configuration. A reproducible model should have low standard deviation.

In academia, reproducibility is not an important factor. Metrics are usually reported as the mean of the metrics from several runs using the same configuration, but the standard deviation is not usually compared.

\subsection{Stability}
We use a distributed training approach for our model, where numerical instability can cause the model to fail. This requires us to restart from a previous checkpoint or abandon the run entirely leading to wasted computational resources. 
Any new feature interaction incorporated into the model should not affect its stability.

Since we continuously retrain our models on newer data, stability across different data distributions is important. Academia usually uses a fixed benchmark dataset to report metrics so stability isn't usually tracked.

\section{Experiments}
\subsection{Metrics}
We use the following metrics to evaluate our models.
\begin{itemize}
    \item HIT@3/save metric\cite{transact}: Although our model predicts multiple actions, we use the most important action \textit{save} to perform offline evaluation of our models. We sort the logged results from each user session according to new predictions and calculate the number of saves in top 3 Pins. We look at this metric compared to the baseline model and report it as a percentage gain or loss.
    \item Memory: This is the peak allocated memory during training as a percentage of the total available GPU memory.
    \item Latency: This is reported as the increase or decrease in the model inference time across a large number of batches compared to the baseline model.
\end{itemize}

\subsection{Learning Experiments}
In this section we present a set of learning experiments we ran to determine which architecture changes are beneficial to the model. The results from these learning experiments are in Table \ref{tab:learning}

\begin{table}
  \caption{Offline Evaluation of Learning Experiments}
  \label{tab:learning}
  \begin{tabular}{ccccc}
    \toprule
    Layer & nlayer/rank & HIT@3/save & Memory & Latency \\
    \midrule
    DCNv2 & 5 & 0.14\% & 69.8\% & +5.8\% \\
    Stacked & 6 & 0.29\% & 71.1\% & +11.2\% \\
    & 7 & 0.38\% &72.4\% & +17.3\% \\
    & 8 & 0.58\% & 73.7\% & +22.8\% \\
    \cmidrule(lr){1-5}
    4x DCNv2 & 2 & 0.05\% & 73.5\% & +21.7\% \\
    Parallel & 3 & 0.33\% & 78.8\% & +44.5\% \\
    \cmidrule(lr){1-5}
    LR DCNv2 & 512 & 0.20\% & 65.4\% & +0.4\% \\
    with ReLU & 1024 & 0.52\% & 65.9\% & +1.2\% \\
     
  \bottomrule
\end{tabular}
\end{table}

\subsubsection{Order of interactions}
We evaluated the model's response to increasing the order of interactions by stacking additional feature interaction layers. In our experiments, we successfully used up to 8 stacked DCNv2 layers without encountering out-of-memory (OOM) issues and observed metric improvements. This tells us that our model benefits from even higher order interactions.

\subsubsection{Parallel interactions}
We checked the model's response to parallel layers designed to learn similar interactions. For instance, parallel DCN layers may individually capture different interaction patterns. In our experiments, running multiple stacked DCNv2 layers in parallel improved model performance. This tells us that our model benefits from learning more than one feature interaction of the same order.

\subsubsection{Non-linearity of interactions}
DCNv2 does not incorporate non-linearity in its architecture. One way to add non-linearity inside DCNv2 is to use the low rank (LR) version with a non-linearity between the two low rank fully connected layers. We replace the full rank DCNv2 layers with low rank layers as the baseline and use ReLU layers as the non-linearity for experimentation. We compared the models with different rank values and observed that our model improved with the inclusion of non-linear interactions.


\subsection{Variants}
We selected several feature interaction architectures that claim better performance than DCNv2 for experimentation. We provide implementation details of the architectures below and also discuss how they fit into our constraints. We also mention how the results from the learning experiments guided us into the final model selection. The results from these experiments are in Table \ref{tab:variants}. A summary of how each of the different architectures performs with respect to the different constraints is in Table \ref{tab:summary}.

\begin{table*}[t]
\centering
  \caption{Offline Evaluation of Variants}
  \label{tab:variants}
  \begin{tabular}{cc}
    \toprule

    Transformer &\begin{tabular}{cccccc}
    nhead & D & nlayer & HIT@3/save & Memory & Latency \\
    \midrule
    2 & 64 & 1 & -0.90\% & 62.2\% & -20.7\% \\
    2 & 64 & 2 & -0.83\% & 67.7\% & -17.7\% \\
    2 & 128 & 2 & -0.47\% & 65.5\% & -16.6\%\\
    8 & 256 & 1 & -0.12\% & 78.6\% & -8.8\%\\
    
\end{tabular} \\
    \midrule
 FinalMLP &\begin{tabular}{cccccc}
    nhead & D & nlayer & HIT@3/save & Memory & Latency \\
    \midrule
    32 & 512 & 4 & -1.30\% & 80.6\% & -3.3\% \\
    32 & 256 & 2 & -1.51\% & 65.7\% & -9.3\% \\
    16 & 512 & 2 & -1.52\% & 75.4\% & -6.6\%\\
    16 & 256 & 4 & -1.67\% & 71.5\% & -5.9\%\\
    
\end{tabular} \\
    \midrule
    GDCN & \begin{tabular}{cccc}
    nlayer & HIT@3/save & Memory & Latency \\
    \midrule
    3 & 0.00\% & 73.3\% & +11.2\% \\
    4 & 0.00\% & 76.7\% & +22.2\% \\
    5 & -0.01\% & 80.0\% & +34.1\% \\
    6 & -0.15\% & 83.4\% & +46.1\% \\

\end{tabular} \\
    \midrule
    DeepLight & \begin{tabular}{cccc}
    hidden\_sizes & HIT@3/save & Memory & Latency \\
    \midrule
    4096x4 &-4.14\% & 65.7\% & -19.3\% \\
\end{tabular} \\
    \midrule
    SDCNv3 & \begin{tabular}{cccccc}
    nDeepCrossLayers & nShallowCrossLayers & HIT@3/save & Memory & Latency \\
    \midrule
    4 & 4 & -6.23\% & 72.2\% & -13.9\% \\
    2 & 2 & -1.21\% & 71.4\% & -25.6\% \\
\end{tabular} \\
    \midrule
    MaskNet & \begin{tabular}{ccccc}
    Type & nlayer & HIT@3/save & Memory & Latency \\
    \midrule
    Stacked & 1 & +0.01\% & 69.2\% & +6.8\% \\
    & 2 & +0.03\% & 76.2\% & +31.9\% \\
    & 3 & +0.01\% & 83.3\% & +57.2\% \\
    Parallel & 2 & +0.13\% & 69.4\% & +17.7\% \\
    & 3 & +0.28\% & 74.0\% & +36.1\% \\
    & 4 & +0.32\% & 78.7\% & +57.5\% \\
\end{tabular} \\
    \midrule
    DHEN & \begin{tabular}{cccc}
    Configuration & HIT@3/save & Memory & Latency \\
    \midrule
    \text{[[MLP, Transformer], [MLP, Transformer]]} & -0.46\% & 75.8\% & -4.4\% \\
    \text{[[DCNv2], [Transformer]]} & 0.63\% & 80.3\% & -5.7\% \\
    \text{[[DCNv2], [MaskNet]]} & 0.77\% & 81.1\% & -1.4\% \\
    \text{[[DCNv2, Masknet]]} & 0.66\% & 82.2\% & -0.7\% \\
\end{tabular} \\
  \bottomrule
\end{tabular}
\end{table*}

\begin{table*}[t!]
    \centering
    \captionsetup{justification=centering}
    \caption{Summary of how architectures perform with respect to different constraints. \\ \cmark \ indicates that the change in this dimension is acceptable given the metric movement. \xmark \ indicates otherwise.}
    
    \begin{tabular}{|c|c|c|c|c|c|}
    \hline
       Variant & Memory & Latency & Hyperparameters & Reproducibility & Stability \\
       \hline
       Transformer  & \xmark & \cmark & \xmark & \cmark & \cmark \\
      \hline
      FinalMLP & \xmark & \cmark & \xmark & \cmark & \cmark \\
      \hline
      GDCN & \xmark & \xmark & \cmark & \cmark & \cmark \\
       \hline
       MaskNet & \cmark & \xmark & \cmark & \cmark & \cmark \\
       \hline
      SDCNv3 & \xmark & \cmark & \xmark & \cmark & \cmark \\
      \hline
      DeepLight & \cmark & \cmark & \xmark & \xmark & \cmark \\
      \hline
       DHEN & \xmark & \cmark & \xmark & \xmark & \xmark \\
       \hline
    \end{tabular}
    
    \label{tab:summary}
\end{table*}

\subsubsection{Transformer}
To use a transformer for feature interaction, we first project our $S$ sparse features into a common dimension $D$.
We grouped our dense features and projected them $C$ times into dimension $D$. We projected the embedding features into dimension $D$. Additionally, we projected the output from our user sequence transformers $U$ times into dimension $D$. This resulted in a final feature set comprising $(S+C+U)$ tokens, each of dimension $D$. We then applied multiple transformer encoder layers for feature interaction. The transformer output tokens were concatenated and processed using an MLP.


In our experiments we set $C=4$ and $U=4$. We used different values of number of heads, token dimension and number of layers as shown in Table \ref{tab:variants}. Transformers are memory hungry, so we were only able to train a 2 layer transformer encoder model without facing OOM errors. This limits the order of feature interactions to 2 which is bad because our model prefers higher order feature interactions. We cannot reduce the batch size to accommodate more layers because engagement metrics are not high enough compared to the baseline. The latency and number of hyperparameters also dissuaded us from exploring this architecture further.

\subsubsection{FinalMLP}
We conducted experiments with a focus on tuning hyperparameters such as the number of layers, the sizes of the hidden layers and the number of heads i.e \textit{k}. We ran into OOM errors with smaller values of $k$. Although our model prefers parallel interactions, we were not able to get the same performance as the baseline using FinalMLP. As we increased the latent dimensions, both memory usage and latency increased  without any performance improvement over the baseline. This led us to conclude that the baseline method was significantly better at learning feature interactions than the multi-head bi-linear fusion technique.

\subsubsection{GDCN}
The only hyperparameter we have to tune for GDCN is the number of layers. We use roughly double the number of parameters in each GDCN layer compared to DCNv2. This limits the order of feature interactions we can learn because of the additional memory used by the parameters. We know our model benefits from much higher order feature interactions, so the gating provided by GDCN is not useful.

\subsubsection{Masknet}
We experimented with two configurations: stacking MaskNet layers sequentially and running them in parallel. Key hyperparameters to tune include the projection ratio, the number of blocks, and the output dimension. In our experiments we use projection ratio = 2.0 and output dimension = 512 and only tuned the number of blocks. We know that our model benefits from parallel feature interactions and non-linearity in the interactions. Both combined, the parallel MaskNet layer outperforms the baseline. This comes with 3 hyperparameters to tune instead of one but that it is not very limiting. We increase the memory consumption and latency quite a bit with this architecture but the increase in model quality is worth it.

\subsubsection{SDCNv3}
We swapped the 4 x stacked DCNv2 layer in our feature interaction layer with stacked SDCNv3 layers which claim better performance. We stacked d x d/2 feature cross layers and experimented with changing the number of layers. We notice that while the SDCNv3 has improved latency it doesn’t converge well on our dataset and thereby drops the offline eval for saves@3. We believe some of these feature interaction layers like SDCNv3 require extensive hyperparameter tuning to make the model perform similar to our baseline model, so we didn't explore it further.

\subsubsection{DeepLight}
DeepLight uses a lightweight interaction layer. To make the comparison fair, we scale up the MLP component of the DeepLight model. Despite this, the model can't learn as useful representations as the baseline. Additionally, we note that the latency numbers reported for this model are before the specified pruning process in their paper. We did not tune all the required hyper parameters for their specified pruning process as the unpruned model performance is far below our baseline.


\begin{table}
  \caption{Online Evaluation of the Launch Candidate}
  \label{tab:online}
  \begin{tabular}{cc}
    \toprule
    Online Metric & Change\\
    \midrule
    Homefeed Save Volume & +1.42\% \\
    Overall Time Spent & +0.39\% \\
  \bottomrule
\end{tabular}
\end{table}

\subsubsection{DHEN}
Since our model prefers parallel interactions and higher-order interactions, we tried combining multiple feature interaction layers using the DHEN architecture. Given the large number of hyperparameters involved, an extensive search is infeasible within our compute budget. The hyperparameters we need to tune include the number of layers, the specific interactions used in each layer, the output size of each layer, and input transformations similar to those used in the transformer. Table \ref{tab:variants} shows some of the variants we tried. The configuration is presented as a list of lists where the first list represents each layer and the second list represents the interactions within the layer.

We note that DCNv2 layers combined with the parallel MaskNet layers have good engagement metric gain. But the DHEN framework is bulky because it involves splitting and concatenating between each layer in the model. This leads to unnecessary GPU memory consumption and latency. We also ran into model stability issues which dissuaded us from pursuing this architecture further.

\subsection{Results}
From Table \ref{tab:summary} we can see that running multiple MaskNet layers in parallel achieves a good trade-off for all of the constraints except for latency. To reduce the latency of the model, we stopped concatenating the input and output of the feature interaction layer before passing it to the MLP layer. We also reduced the size of the hidden layers in the MLP. This resulted in a slight reduction in metric gain and an overall reduction in the latency compared to the baseline. Since we know that the model prefers higher order feature interactions, we stacked 4 DCNv2 layers on top of the parallel MaskNet layers. We tuned the hyperparameters of the MaskNet model to get no latency increase and net zero increase in the number of parameters while increasing the memory consumption by only 5\% absolute allowing us to use the same batch size. We also ran online A/B tests using this new architecture, the results of which are in Table \ref{tab:online}. The final configuration has 3 parallel MaskNet blocks with a projection ratio of 2.0 and output dimension of 512 and is shown in Figure \ref{fig:masknet}.
We confirmed that this architecture doesn't decrease reproducibility and did not observe any model instability issues in any of our runs.

\section{Conclusion}
In this paper, we presented several constraints that industrial recommendation systems have to deal with and detailed how they influenced our selection of a new feature interaction architecture. This change improved our model and has been successfully deployed in the Homefeed recommendation system at Pinterest. We continue iterating on our feature interaction layer to further improve our model. We hope these findings can inspire collaboration between academia and industry to propose new architectures that work well in industrial constraints.

\section{Acknowledgements}
The authors would like to thank Dylan Wang, XianXing Zhang, Nikil Pancha, Meng Qi, Liang Zhang and others who contributed, supported and collaborated with us.

\bibliographystyle{ACM-Reference-Format}
\bibliography{main}

\end{document}